\documentclass[twocolumn,prl,aps,superscriptaddress,showpacs]{revtex4}

\usepackage{bm}
\usepackage{graphicx}
\usepackage{color}

\begin{document}

\title{Pressure-phase diagram of UCoGe by ac-susceptibility and resistivity measurements}

\author{E Hassinger, D Aoki, G Knebel and J Flouquet}

\affiliation{CEA Grenoble, 17 rue de Martyrs, 38000 Grenoble, France}


\begin{abstract}
UCoGe is one of the few compounds showing the coexistence of ferromagnetism and superconductivity at ambient pressure. With $T_{Curie}=3$\,K and $T_{SC}=0.6$\,K it is near a quantum phase transition; the pressure needed to suppress the magnetism is slightly higher than 1\,GPa. We report simultaneous resistivity and ac-susceptibility measurements under pressure on a polycrystal with very large single-crystalline domains and a resistivity ratio of about 6.
Both methods confirm the phase diagram established before by resistivity measurements on a polycrystal. The ferromagnetic phase is suppressed for $P \approx 1.2$\,GPa. Astonishingly, the superconductivity persists at pressures up to at least 2.4\,GPa. In other superconducting and ferromagnetic heavy fermion compounds like UGe$_2$ and URhGe, the superconducting state is situated only inside the larger ferromagnetic region. Therefore, UCoGe seems to be the first example where superconductivity extends from the ferromagnetic to the paramagnetic region.
\end{abstract}
\maketitle
\section{Introduction}
The coexistence of ferromagnetism (FM) and superconductivity (SC) is an intriguing problem ever since the microscopic understanding of superconductivity \cite{Matthias1958a, Ginzburg1957}. In a conventional superconductor, a magnetic field destroys the superconducting state by breaking the Cooper pairs which are in a spin singlet state. However, one can imagine that a Cooper pair with a total spin of one (spin triplet) can exist even in the strong internal field of a ferromagnet. The question is, if the superconductivity in these compounds is mediated by magnetic fluctuations due to the proximity to the critical point where the suppression of the ferromagnetic phase occurs. The pressure-temperature phase diagram of a ferromagnetic superconductor near the critical pressure, has been studied theoretically \cite{Roussev2001,Fay1980}, but until the discovery of UCoGe in 2007 \cite{Huy2007}, only one other ferromagnetic superconductor, UGe$_2$, had a ferromagnetic critical point in an attainable pressure region \cite{Saxena2000}. In UGe$_2$, the superconducting phase lies entirely within the ferromagnetic phase. In URhGe, increasing the pressure drives the system away from the quantum phase transition. In UCoGe, resistivity measurements on a polycrystal have shown, that the superconducting phase persists to the highest measured pressure which is higher than the pressure, where ferromagnetism is suppressed \cite{Hassinger2008}. As resistivity can indicate also non-bulk superconductivity we decided to do simultaneous ac-susceptibility and resistivity measurements, in order to check previous measurements. With a polycrystal with very large grains (almost single crystalline), we confirm the previously published phase diagram.
\section{Experimental methods}
The almost single-crystalline sample is Czochralski pulled out of a melted mixture of the ingredients in a 1:1:1 ratio and annealed (see \cite{Hassinger2008}) and has a residual resistivity ratio (RRR) of around 6, lower than the best polycrystals ($\approx 25$).
Two coils of 20 turns each are wound around the sample and the 4 wires for resistivity measurements are spot welded at the free end of the same sample. Pressure was created by a CuBe/NiCrAl hybrid pressure cell with Daphne oil as pressure transmitting medium. The pressure was measured via the pressure dependent superconducting transition temperature of a piece of lead in the pressure chamber. Typical measurement frequencies were 331 Hz for the ac-susceptibility measurements. We adjusted the phase so that the ferromagnetic transition was totally in the real part of the signal. Further, susceptibility measurements were also performed in a diamond anvil cell with a detection coil inside the pressure chamber \cite{Braithwaite2007}. 
\section{Results and discussion}
\begin{figure}[h!]%
\begin{minipage}[b]{17pc}
\includegraphics[width=17pc]{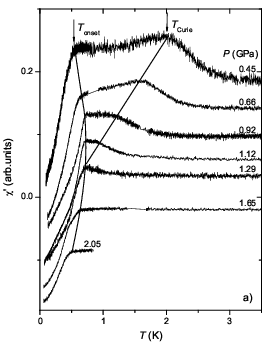}
\end{minipage}\hspace{4pc}
\begin{minipage}[b]{17pc}
\includegraphics[width=17pc]{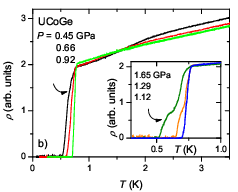}
\end{minipage}
\caption{\label{susc+rho}a) Real part of ac-susceptibility at different pressures. b) Resistivity at the same pressures.}
\end{figure}
Figure \ref{susc+rho} shows the results of ac-susceptibility and resistivity measurements at 6 different pressures. In the susceptibility (figure \ref{susc+rho}a), the ferromagnetic transition is clearly visible as a maximum. The transition temperature $T_{Curie}$ was defined as the temperature of the maximum. With pressure, the maximum shifts to lower temperatures and is strongly attenuated. The superconducting temperature $T_{SC}$, defined as the onset temperature of the shielding signal, increases with pressure up to around 1.1\,GPa and then decreases again.At 1.65\,GPa, there is no sign of a ferromagnetic transition, i.e. a completely flat behavior, characteristic of a material far from a FM instability, is restored above $T_{SC}$.

In the resistivity (figure \ref{susc+rho}b), the anomaly at $T_{Curie}$ is rather broad and we do not observe a well defined feature. This may be due to the rather low sample quality with RRR $\approx 6$. This broad anomaly shifts to lower temperatures and is also attenuated. The superconducting transition width goes through a clear sharp minimum at $\approx 1.1$\,GPa. 
The onset temperature of the superconducting transition does not change with pressure, only $T(\rho=0)$ changes. Below $P=1.2$\,GPa $T(\rho=0)$ corresponds exactly to the $T_{SC}$defined as above from susceptibility. Above this pressure, $T(\rho=0)$ is slightly lower. Interestingly, above this pressure also a double step behavior develops, which is not seen at low pressures.

\begin{figure}[h]
\includegraphics[width=20pc]{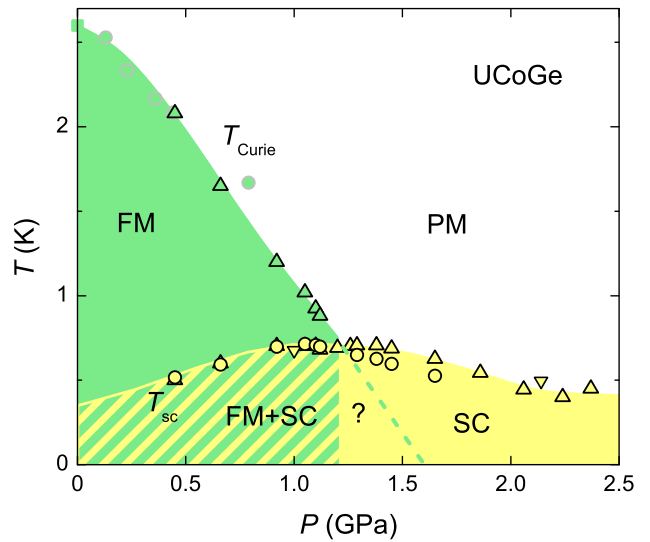}
\caption{\label{phasediag}Pressure-temperature phase diagram of a polycrystalline sample of UCoGe by susceptibility (upward triangles: piston cylinder cell, downward triangles: diamond anvil cell) and resistivity (circles) measurements. The grey circles are from reference \cite{Hassinger2008}.}
\end{figure}

Figure \ref{phasediag} shows the pressure-temperature phase diagram established from these and previous measurements. The qualitative behavior of the phase transitions is the same as in reference \cite{Hassinger2008}. The Curie-temperature $T_{Curie}$ decreases linearly with pressure, but the slope is steeper than the one of the previous phase diagram. The superconducting phase is dome-like with the maximum at the pressure where the ferromagnetic transistion line meets the superconducting phase boundary. It extends up to the highest measured pressure. 

The phase diagram presented here is different to all theoretical predictions, and to the phase diagrams of other ferromagnetic superconductors, namely UGe$_2$, where the superconducting phase lies entirely within the ferromagnetic region and no superconducting phase appears in the paramagnetic (PM) regime. But in UGe$_2$, SC is directly linked to the duality between its two competing ferromagnetic states (FM1 and FM2) \cite{Pfleiderer2002}. From the presented data however, it is clear that in UCoGe SC appears also in the PM state. An extrapolation of $T_{Curie}$ to 0\,K would give a quantum critical point (QCP) close to 1.6\,GPa. Up to now, the real behaviour in the pressure region from $P(T_{Curie}=T_{SC})$ up to $P(T_{Curie}=0)$ is not clear.

\begin{figure}[h]
\includegraphics[width=21pc]{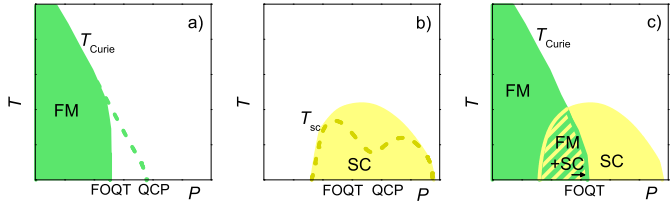}
\caption{\label{schema}Behavior of superconductivity and ferromagnetic phases at a first order (solid line) or second order (dashed line) phase transition.}
\end{figure}

Figure \ref{schema}a represents in the filled area the hypothetical variation of FM in a first order quantum transition (FOQT) \cite{Belitz1999} neglecting any feedback of SC; the dashed line the hypothetical extrapolation of a second order quantum critical point (QCP). For clarity, the difference between the FOQT and the QCP has been exagerated. In figure \ref{schema}b is shown the superconducting dome centered near the FOQT (filled area) assuming that the first order nature of FM wipes out the SC minimum at the QCP (dashed line \cite{Fay1980,Roussev2001}) as experimentally observed. Figure \ref{schema}c represents what may be the result of the interplay between FM and SC including a feedback between the two phases: In a narrow regime one may go from the PM to the simple SC phase and then to the FM+SC coexisting phase. The situation will be rather similar to the one found in the antiferromagnetic (1,1,5) Ce compound CeRhIn$_5$ under pressure \cite{Knebel2006, Yashima2007}. However, neither resistivity nor susceptibility are sensitive to this transition within the SC phase.
\section{Conclusion}Simultaneous ac-susceptibility and resistivity measurements on one UCoGe crystal confirm the previously published pressure-temperature phase diagram. However, only specific heat measurements will be able to decide, if the superconducting transition is bulk in the whole pressure range. If this is the case, UCoGe presents a completely different phase diagram compared to other ferromagnetic superconductors and is therefore a unique case to study the interplay between ferromagnetism and superconductivity at the critical point.
\section*{Acknowledgements}Financial support has been given by the French ANR within the programs ECCE and NEMSICOM. Thanks to D. Braithwaite for help with the DAC setup.
\providecommand{\newblock}{}

\end{document}